\documentclass[showpacs,aps,amsmath]{revtex4}
\usepackage{graphicx}
\usepackage{epsfig}

\def\text#1{\mbox{#1}}
\def\equ#1{(\ref{#1})}

\def\be#1{\begin{equation}\label{#1}}
\def\ee{\end{equation}}
\def\equ#1{(\ref{#1})}

\begin{document}

\title{
  One-dimensional Dirac oscillator in a thermal bath
}

\author{
  M. H. Pacheco,
  R. R. Landim\footnote{e-mail: renan@fisica.ufc.br}, and
  C. A. S. Almeida\footnote{e-mail: carlos@fisica.ufc.br}
}

\affiliation{
  Departamento de F\'{\i}sica, Universidade Federal do Cear\'{a},\\
  Caixa Postal 6030, 60455-760 Fortaleza, Cear\'{a}, Brazil
}

\begin{abstract}
  We analyze the one-dimensional Dirac oscillator in a thermal bath.  We found that
  the heat capacity is two times greater than the heat capacity of the one-dimensional
   harmonic oscillator for higher temperatures.
\end{abstract}

\pacs{03.65.Pm; 03.65.Ge; 11.10.Qr; 11.10.Wx}

\maketitle
The name Dirac oscillator was first introduced by Moshinky and
Szczepaniak \cite{szczep} for a Dirac equation in which the
momentum $\mathbf{p}$ is replaced by $\mathbf{p\rightarrow
p-}\mathit{i}$m$\mathbf{\omega \beta r}$ with $\mathbf{r}$ being
the position vector, m the mass of the particle and $\omega$ the
frequency of the oscillator. The name was given because the
determination of the spectrum and eigenstate of a Dirac oscillator
\cite{szczep} was obtained from that of an ordinary oscillator
with a very strong spin-orbit coupling term. The concept, though
not the name, had been discussed previously first by Ito et al.
\cite{ito} and later by Cook \cite{cook} and Ui and Takeda
\cite{Ui:zg} and possibly others. Besides intrinsic mathematical
interest the study of the Dirac oscillator has invoked much
attention because of its various physical applications. For
instance, the last three chapter of a book of Moshinsky and
Smirnov \cite{mosh} deal with the Dirac oscillator relativistic
interactions between systems of one, two and three particles.
Moreno and Zentella showed \cite{moreno} that it could be the
object of an exact Foldy-Wouthuysen transformation. Benitez et al.
\cite{Benitez:te} found the electromagnetic potential associated
with the Dirac oscillator, and showed that this exactly soluble
problem has a hidden supersymmetry, which is responsible for the
special properties of its energy spectrum. They also calculated
the related superpotential and discussed the implications of this
supersymmetry on the stability of the Dirac sea.

Rozmej and Arvieu~\cite{Rozmej:1999jv} have shown a very
interesting analogy between the relativistic Dirac oscillator and
the Jaynes-Cummings model. They showed that the strong spin-orbit
coupling of the Dirac oscillator produces the entanglement of the
spin with the orbital motion similar to what is observed in the
model of quantum optics.

More recently Nogami and Toyama \cite{nogami} and Toyama et al
\cite{toyama} have studied the behaviour of wave packets of the
Dirac oscillator in the Dirac representation in (1+1) dimensions.
The aim of these authors was to study wave packets which could
possibly be coherent. This reduction of the dimension was brought
as an attempt to get rid of spin effects and to concentrate on the
relativistic effects.

As a matter of fact, the one-dimensional relativistic Dirac
equation has been largely used to treat physical problems where
relativistic effects could play an important role. Indeed, this
treatment gives enlightening information about more realistic
models in solid state physics \cite{solid state}, particularly in
semiconductor theories \cite{Renan:1999tr}.

In spite of the great number of papers that has been recently
published concerning the solution and properties of the Dirac
oscillator, as far as we know no one has reported on its
thermodynamics properties. In order to overcome this lack of
information, in this letter we study the one-dimensional Dirac
oscillator in a thermal bath.

Let us consider  the energy spectrum for a one-dimensional Dirac
oscillator~\cite{adame}

\begin{equation}
E_n=\sqrt{{[2(n+1)]\hbar\omega mc^ 2+m^{2}c^4}}, \label{energy}
\end{equation}
where $m$ is the mass of the particle, $\omega$ is the angular
frequency of the oscillator and $n$ is a positive integer number.
As we can see, for $n+1<< mc^2/2\hbar\omega$,  the spectrum of the
one-dimensional Dirac oscillator is approximated, up to $E_0$,  to
the spectrum of the one-dimensional harmonic oscillator.

The partition function of the Dirac oscillator at temperature $T$
is obtained through the Boltzmann factor,

\begin{equation}
Z=\sum_{n=0}^{\infty}e^{-(E_n-E_0)\beta}=e^{\sqrt{b}\beta}\sum_{n=0}^{\infty}
e^{-\sqrt{an+b}\beta}, \label{Z} \end{equation}
were $\beta=1/kT$,
$k$ the Boltzmann constant, $a=2 \hbar\omega mc^2$,
$b=m^2c^4+2\hbar\omega mc^2$, and we are using a frame in the heat
bath. We have subtracted the ground state energy in order to make
a comparison with the non-relativistic harmonic oscillator. All
thermodynamics quantities for the Dirac oscillator are obtained
through the  partition function \equ{Z}.

Let us first analyze the convergence of the partition function.
The function $f(x)=\exp(-\sqrt{ax+b}\beta)$ is a monotonically
decreasing function and the corresponding integral \be{int}
I(\beta)=\int_{0}^{\infty}
e^{-\sqrt{ax+b}\beta}dx=\frac{2}{a\beta^{2}}
e^{-\sqrt{b}\beta}(1+\beta\sqrt{b}),
 \ee
is convergent. Thus from the theorems of convergent series, this
implies that the partition function is convergent.

In order to evaluate the partition function, we  can use  the
Euler-MacLaurin formula

\begin{equation}
\sum_{n=0}^\infty f(n)=\frac{1}{2}f(0)+\int_0^\infty
f(x)dx-\sum_{p=1}^\infty\frac{1}{(2p)!}B_{2p}f^{(2p-1)}(0),
\label{em}
\end{equation}
where $B_{2p}$ are the Bernoulli numbers, $B_2=1/6,
B_4=-1/30,...$. They are defined through the series

\begin{equation}
\frac{t}{e^t-1}=\sum_{n=0}^\infty B_n\, \frac{t^n}{n!}.
\label{ber}
\end{equation}

Then the partition function can be written as

\be{partf}
Z=\frac{1}{2}+\frac{2}{a\beta^{2}}(1+\beta\sqrt{b})-e^{\sqrt{b}\beta}\sum_{p=1}^\infty\frac{B_{2p}}{(2p)!}f^{(2p-1)}(0).
\ee

To compute the partition function, we need to calculate the sum in
the above expression. For our case it  can be done only by
numerical methods. Up to $p=2$ this sum can be written as:
\begin{eqnarray}
e^{\sqrt{b}\beta}\sum_{p=1}^2\frac{B_{2p}}{(2p)!}f^{(2p-1)}(0)=
\nonumber
 \\  \left({\frac {1}{24}} \,{\displaystyle \frac {\beta
\,a\,}{\sqrt{b}}}- {\displaystyle \frac {1}{1920}} \,
{\displaystyle \frac {\beta \,a ^{3}\,}{b^{5/2}}}  -
{\displaystyle \frac {1}{1920}} \,{\displaystyle \frac {\beta
^{2}\,a^{3}\,}{b^{2}}}  - {\displaystyle \frac {1}{5760} }
\,{\displaystyle \frac {\beta ^{3}\,a^{3}\,}{b^{3/2}}}\right).
\label{sum}
\end{eqnarray}

Before computing numerically the partition function, let us
analyze the case of  higher temperatures. This correspond to
$\beta\ll 1$. All terms in the sum of \equ{partf} have a positive
power in $\beta$, which are very small compared with the term
$2(1+\beta\sqrt{b})/a\beta^2$. Hence, for $\beta\ll 1$
\begin{equation}
Z\simeq\frac{2}{a\beta^{2}}, \label{higgerT}
\end{equation}
Now we can easily obtain the mean energy and the heat capacity of
the Dirac oscillator for higher temperatures
\begin{equation}
U=-\frac{\partial \ln Z}{\partial \beta} \simeq 2kT,\label{em1}
\end{equation}
\begin{equation}
C\simeq \frac{\partial U}{\partial T}\simeq 2k. \label{cv}
\end{equation}
These results show that, for higher temperatures, the mean energy
and the heat capacity for the Dirac oscillator is two times the
mean energy and the heat capacity of the non-relativistic one
dimensional harmonic oscillator. This is due, only to relativistic
effects, since in one dimension there is no spin coupling.

\begin{figure}
\centerline{
\epsfig{figure=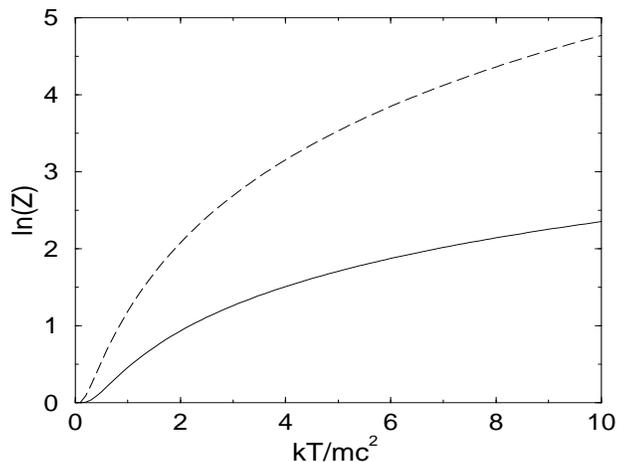,width=230pt,height=180pt} }
\caption[caption]{The free energy for the harmonic and the Dirac
oscillator for higher temperatures and $\hbar\omega/mc^2=1$.
Dashed line corresponds to the Dirac oscillator and the solid line
the harmonic oscillator.} \label{FIG:1.1}
\end{figure}

Now we briefly discuss our numerical results on the calculation of
the partition function \equ{partf}. First of all, we should
mention that, using \equ{partf}, the curves for the three thermal
functions, namely, mean energy, heat capacity and free energy, are
identical to those obtained for the nonrelativistic
one-dimensional harmonic oscillator, in the limit of low
temperatures ($0$ to $400K$).

\begin{figure}
\centerline{
\epsfig{figure=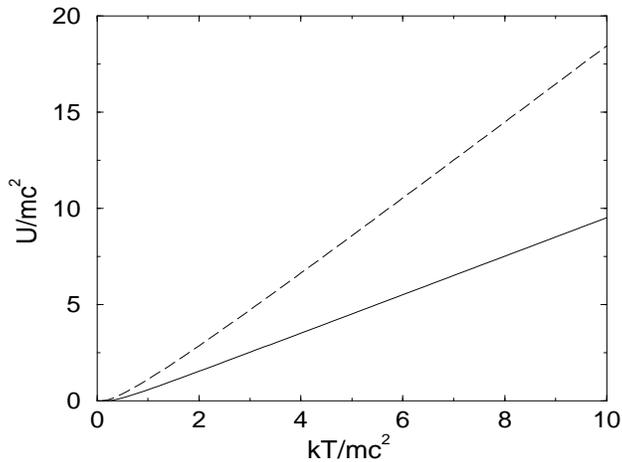,width=230pt,height=180pt} }
\caption[caption]{The mean energy for the harmonic and the Dirac
oscillator for higher  temperatures and $\hbar\omega/mc^2=1$.
Dashed line corresponds to the Dirac oscillator and the solid line
the harmonic oscillator.} \label{FIG:1.2}
\end{figure}

\begin{figure}
\centerline{
\epsfig{figure=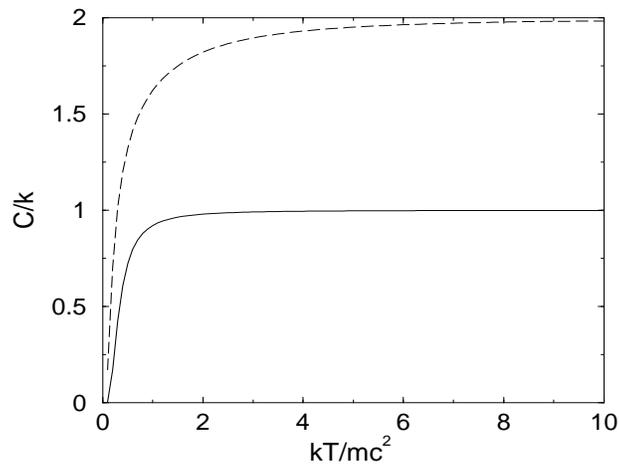,width=230pt,height=180pt} }
\caption[caption]{The heat capacity for the harmonic and the Dirac
oscillator for higher temperatures and $\hbar\omega/mc^2=1$.
Dashed line corresponds to the Dirac oscillator and the solid line
the harmonic oscillator.} \label{FIG:1.3}
\end{figure}

We display a comparison between the Dirac oscillator and the
harmonic oscillator only, for high temperatures
(Fig.~\ref{FIG:1.1} and Fig.~\ref{FIG:1.2}). It is seen that the
free energy and the mean energy are greater for the Dirac
oscillator than for the harmonic oscillator. From
Fig.~\ref{FIG:1.3} it is also seen that the heat capacity for the
Dirac oscillator is two times the heat capacity for the harmonic
oscillator, result that was anticipated by the analytical
calculations presented above.

It is worthwhile to mention that numerical calculations of the
partition function for high temperatures ($\beta\ll 1$), force us
to handle several very small variables, which could lead to
deceiving results. So, it is required a precise estimation of the
involved physical quantities. Therefore, we choose the angular
frequency of the oscillator as $\omega=10^{20}Hz$ in the region of
high temperatures. We used adimensional quantities in the figures.
The temperature ranges from $10^{8}K$ to $10^{10}K$ ($0.01$ to
$1.0$ $MeV$).

As an extension of this work, we are currently studying a chain of
Dirac oscillators in a thermal bath. Also, construction of higher
dimensional Dirac oscillator in finite temperature, despite its
technical difficulties, would be very interesting and could shed
light on relativistic effects in statistical and solid state
physics.

Some models used the nonrelativistic harmonic oscillator potential
for describing confinement of quarks in mesons and baryons
\cite{Ravndal:sr}. More recently, some authors suggested
\cite{moreno,adame} that the Dirac oscillator could be a good
candidate to be used as the confinement potential in heavy quark
systems. On the other hand, recently ultrarelativistic heavy ion
experiments have search for the quark gluon plasma, a novel phase
of QCD in which quarks and gluons are deconfined
\cite{Braun-Munzinger:2000mv}. We expect that studies on
quark-gluon plasma models and their thermal properties could be
subsidized by our results.

\vspace{0.3in} \centerline{\bf ACKNOWLEDGMENTS} \vspace{0.3in} M.
H. P., R. R. L. and C. A. S. A. were supported in part by Conselho
Nacional de Desenvolvimento Cient\'{\i}fico e
Tecnol\'{o}gico-CNPq.



\vspace{0.3in} \centerline{\bf REFERENCES}

\begin{references}
\bibitem{szczep} M. Moshinsky and A. Szczepaniak, Jour. Phys. 1989 {\bf A22} L817.
\bibitem{ito} D. Ito, K. Mori and E. Carriere, Nuovo Cim. 1967 {\bf 51A} 1119.
\bibitem{cook} P. A. Cook, Lett. Nuovo Cimento 1971 {\bf 1} 419.
\bibitem{Ui:zg}
H.~Ui and G.~Takeda,
Prog.\ Theor.\ Phys.\  {\bf 72}, 266 (1984).
\bibitem{mosh} M. Moshinsky and Y. Smirnov, The harmonic oscillator in modern
physics Hardwood Academic Publishers, The Netherlands 1996, pp.
289-404.
\bibitem{moreno} M. Moreno and A. Zentella,  Jour. Phys. 1989 {\bf A22} L821.
A. L. Salas-Brito, Phys. Rev. Lett. 1990 {\bf 64} 1643.
\bibitem{Benitez:te}
J.~Benitez, R.~P.~Martinez y Romero, H.~N.~Nunez-Yepez and
A.~L.~Salas-Brito,
Phys.\ Rev.\ Lett.\  {\bf 64}, 1643 (1990).
\bibitem{Rozmej:1999jv}
P.~Rozmej and R.~Arvieu,
J.\ Phys.\ A {\bf 32}, 5367 (1999) [arXiv:quant-ph/9903073].
\bibitem{nogami} Y.Nogami  and F. M. Toyama, Can. J. Phys. 1996 {\bf 74} 114.
\bibitem{toyama} F.M. Toyama, Y.  Nogami  and F. A. B. Coutinho, Jour. Phys. 1997 {\bf A30} 2585.
\bibitem{solid state} T. Loucks, Phys. Rev. 1965 {\bf A139} 1333.
\bibitem{Renan:1999tr}
R.~Renan, M.~H.~Pacheco and C.~A.~Almeida,
J.\ Phys.\ A {\bf 33}, L509 (2000) [arXiv:cond-mat/9911356].
\bibitem{adame}F. Dom\'\i nguez-Adame and M. A. Gonz\'alez, Europhys. Lett. 1990 {\bf 13(3)} 193.
\bibitem{Ravndal:sr}
F.~Ravndal,
Phys.\ Lett.\ B {\bf 113}, 57 (1982).
\bibitem{Braun-Munzinger:2000mv}
P.~Braun-Munzinger,
Nucl.\ Phys.\ A {\bf 681}, 119 (2001) [arXiv:nucl-ex/0007021].

\end{references}
\end{document}